\documentclass[12pt]{article}
 \newcommand{\be}[3]{\begin{equation}  \label{#1#2#3}}     
 \newcommand{\bib}[3]{\bibitem{#1#2#3}}
\newcommand{\ee}{ \end{equation}}
\newcommand{\ba}{\begin{array}}
\newcommand{\ea}{\end{array}}

\begin{document}

\thispagestyle{empty}
\rightline{UPR-787-T}
\rightline{HUB-EP-97/93}
\rightline{hep-th/9712221}

\vspace{1.5truecm}

\centerline{\bf \Large The entropy of near-extreme N=2 black holes}
\vspace{2truecm}

\centerline{
  {\bf Klaus Behrndt}$^{a,\ A}$ , {\bf Mirjam Cveti{\v c}}$^{b, \  B}$ 
  \quad and \quad
  {\bf Wafic A.\ Sabra}$^{a,c, \ C}$ 
 \footnote{e-mail: $^A$behrndt@qft2.physik.hu-berlin.de, 
                   $^B$cvetic@cvetic.hep.upenn.edu,
                   $^C$sabra@qft2.physik.hu-berlin.de
  }
 }
\vspace{.5truecm}
\centerline{\em $~^a$Humboldt University Berlin}
\centerline{\em Invalidenstrasse 110, 10115 Berlin, Germany}
\vspace{.3truecm}
\centerline{\em $^b$Department of Physics and Astronomy}
\centerline{\em University of Pennsylvania, Philadelphia, PA 19104-6396, USA}
\vspace{.3truecm}
\centerline{\em $^c$Department of Physics}
\centerline{\em Queen Mary  and Westfield College, London E1 4NS, U.K.}

\vspace{1.2truecm}


\begin{abstract}

\noindent
We give an explicit form of the classical entropy for four-dimensional
static near-BPS-saturated black holes of generic N=2 superstring
vacua.  The expression is obtained by determining the leading
corrections in the non-extremality parameter to the corresponding
BPS-saturated black hole solutions.  These classical results are
quantitatively compared with the microscopic leading order corrections
to the microscopic result of Maldacena Strominger and Witten for N=2
BPS-saturated black holes.
 
\end{abstract}

\bigskip \bigskip

\newpage

%
%
\section{Introduction}

Over the past two years a dramatic progress has been made in
understanding the microscopic origin of the black holes (for a review
see for example \cite{005} and references therein), in particular for
the five-dimensional BPS-saturated solutions \cite{010} of
toroidally compactified string theory, i.e.  vacua with $N=4$ and
$N=8$ supersymmetry.  Soon after the microscopics of
near-BPS-saturated black hole solutions both static
\cite{020} and rotating ones \cite{022}
was addressed using the $D$-brane dynamics, however  the arguments
became more heuristic and less rigorous, in particular
 for four-dimensional BPS- and
near-BPS-saturated solutions.

BPS-saturated solutions of $N=4,8$ vacua possess large symmetry, and 
corrections  to  the 
low-energy effective action can appear only through higher derivative
terms like $R^2$ or $F^4$,  however, there are {\it no loop-corrections}
expected.

On the other hand  BPS-saturated solution of $N=2$  string vacua
can receive corrections  due to the 
quantum (loop) corrections already to the lowest
derivative
terms in the effective action.  This rich structure makes it however
much more difficult to find an explicit form or such corrections to the
BPS-saturated black hole solutions.  Nevertheless, solutions
for general (stationary) BPS-saturated solutions
which couple to arbitrary vector fields and trivial hyper-multiplets
have been found \cite{030}.  These solutions are given in terms
of an algebraic constraint for the symplectic section, which can
be solved explicitly for many interesting cases. The existence-proof of such 
solutions
heavily relies  on the remaining supersymmetry of the solutions,
i.e. they satisfy the Killing spinor equations.  In a recent work of
Maldacena, Strominger and Witten \cite{060}, and Vafa \cite{026}, the
rigorous derivation of the microscopic entropy for a  class of
BPS-saturated black holes \cite{027}, \cite{028} was given. It
accounts for both the tree-level and loop corrections to the entropy
of such  black holes; they are microscopically described 
as a five-brane wrapping around $C_4\times S^1$ of M-theory
compactified on $CY_3\times S^1$. Here $CY_3$ is the Calabi-Yau
three-fold and $C_4$ is the four-cycle of $CY_3$. The additional
electric charges correspond to the momentum along $S^1$ and the
background value of the self-dual three-form of the five-brane.  This
work supersedes more heuristic approaches \cite{029} which
addressed the microscopic origin of these black holes within the (intersecting)
$D$-brane approach.

In view of this recent progress a natural open question is to
address the {\it  non-extreme black hole solutions of $N=2$ vacua} with
the hope that once such classical solutions are understood, the next
step should be to address the microscopics of such solutions,
hopefully with a comparable rigor as the recent work on the
BPS-saturated solutions.~\footnote{In addition, the radiation rates in
such black hole backgrounds could provide another dynamical test to
probe the internal structure of such black holes, as it has already
been done for black holes of $N=4,8$ vacua (for a review see
\cite{140} and references therein).}  While the explicit solutions for
general {\it non-extreme} four-(and five-)-dimensional static
\cite{100}, \cite{110} (\cite{024}) and rotating \cite{120}
(\cite{130}) black holes of $N=4,8$ string vacua have been
constructed, the state of affairs for non-extreme black holes of $N=2$
vacua is still a poorly explored territory.

The work by Kastor and Win \cite{040} constitutes the first attempt to
address such non-extreme static solutions. However, their approach
already reveals difficulties in finding an explicit solution of the
equations of motion for a generic $N=2$ vacuum. They took an
Ansatz which coincides with that of the non-extreme black holes of
toroidally compactified string vacua \cite{100}, \cite{050} and it 
turned out to be a solution only  for a very
restricted class of  $N=2$ string vacua, i.e.  Type II superstring  [M-theory] 
compactified on a  restricted   Calabi-Yau three-fold [restricted Calabi-Yau
three-fold $\times  S^1$], which  can effectively be reduced to that of a
torus.
 
The aim of this paper is to further shed light on non-extreme static
black holes for $N=2$ string vacua. In particular, we concentrate on
{\it near-BPS-saturated static black hole solutions for generic
four-dimensional $N=2$ string vacua}.  The entropy of the
near-BPS-saturated solution is terms of the entropy of the
BPS-saturated solution and corrections in terms of the {\it
non-extremality parameter} which parameterizes a deviation from the
BPS-saturated limit.  This classical result provides a starting point
to address the microscopic origin of these corrections.  Note, that
the ultimate goal of the program is to gain the (macroscopic and
microscopic) understanding of {\it generic non-extreme solutions} both
static and rotating ones, both in four-and five dimensions. 
For a recent discussion for extreme solution or $N=2$, $D=5$
supergravity see e.g.\ \cite{035}. Along these
(ambitious) directions we therefore touch only the tip of the iceberg.

The paper is organized in the following way. In Section 2 we spell out
the action and the constraints of a generic $N=2$ supergravity
theories in four-dimensions ($D=4$) and discuss their BPS-saturated
black hole solutions.  In Section 3 we derive equations of motion for
a generic $N=2$ supergravity and employ them to obtain the leading
corrections (in the non-extremality parameter) to the entropy for the
near-BPS-saturated black holes for generic $N=2$ vacua, i.e. for
TypeII string theory compactified on a generic Calabi-Yau
three-fold. In Section 4 we study  microscopic origin of such
corrections.

\section{$N=2$  Supergravity and BPS-Saturated Black Holes}

We start with the  action and  the equations of motion of $N=2$
supergravity in $D=4$. Considering only gravity and vector multiplets
the low energy action is given by
\be010
S \sim \int \sqrt{-g} \, d^4x \Big[R - 2 g_{A \bar B}
\partial^{\mu} z^A \partial_{\mu} \bar z^{B} - {1 \over 4} 
F^I_{\mu\nu} ({^{\star}G_{I}})^{\mu\nu} \Big]
\ee
with the gauge field $G_{I\, \mu\nu}$ given by
\be020
G_{I\, \mu\nu} = \mbox{Re} {\cal N}_{IJ} F^{J}_{\mu\nu} - 
 \mbox{Im} {\cal N}_{IJ} {^{\star}}F^J_{\mu\nu} \ .
\ee
and $I,J = 0,1 .... n_v$, counts the number of vector multiplets.  The
complex scalar fields $z^{A}$ ($A=1..n_{v}$) parameterize a special
K\"ahler manifold with the metric $g_{A\bar{B}} = \partial_{A}
\partial_{\bar{B}} K(z,\bar{z})$, where $K(z,\bar{z})$ is the K\"ahler
potential.  Both, the gauge field coupling as well as the K\"ahler
potential are expressed in terms of the holomorphic prepotential $F(X)$
\be030 
\ba{l} 
e^{-K} = i (\bar{X}^I F_I - X^I
\bar{F}_{I}) \\ {\cal N}_{IJ} = \bar{F}_{IJ} + 2i {({\rm Im} F_{IL} X^{L}) 
({\rm Im} F_{JM} X^{M}) \over {\rm Im} F_{MN} X^{M} X^{N}}\ , 
\ea \ee 
with $F_{I} = {\partial F(X) \over \partial X^{I}}$ and $F_{MN} =
{\partial^{2} F(X)\over \partial X^{M} \partial X^{N}}$. The scalar
fields $z^{A}$ are defined by
\be040 
z^{A} = {X^{A} \over X^{0}} \ .
\ee 
The form of the   static BPS-saturated solutions
 is  of the form \cite{030},
\cite{032}:
\be050
\ba{l}
ds^2 = - e^{2U} dt^2 + e^{-2U} dx^m dx^m \quad , \quad e^{-2U} = e^{-K}\ , \\
F^I_{mn} = \epsilon_{mnp} \partial_p H^I \quad , \quad
G_{I\, mn} = \epsilon_{mnp} \partial_p H_I\ ,
\ea \ee
where $(H^I, H_I)$ are harmonic functions and the $X^I$ are determined by
\be060
i(X^I - \bar X^I) = H^I \qquad , \qquad i(F_I - \bar F_I) = H_I \ .
\ee
We have  assumed  that $e^{-2U}=e^{-K}\to 1$  for $r\to \infty$, i.e. we 
have fixed the  K\"ahler gauge.

For the sake of concreteness we will consider the model defined by
the prepotential
\be070
F(X) = {d_{ABC} X^A X^B X^C \over X^0} \ .
\label{prep}\ee

If all harmonic functions are non-trivial, one cannot solve
(\ref{060}) explicitly, but as long as $H^0 = 0$ the complete
solution to prepotential (\ref{070}) is given by \cite{028}
\be072
X^0 = \sqrt{ d_{ABC}H^A H^B H^C \over H_0 + {1 \over 12} D^{AB} H_A H_B}
\qquad ,  \qquad X^A = {1 \over 6} X^0 D^{AB} H_B - {i\over 2} H^A \ , 
\ee
where $D^{AB} = (d_{ABC} H^C)^{-1}$. Inserting this into (\ref{050})
one finds for the metric
\be074
e^{-2U} = \sqrt{ \tilde H_0 d_{ABC}H^A H^B H^C} \qquad , \qquad 
\tilde H_0 =  H_0 + {1 \over 12} D^{AB} H_A H_B\ .
\ee
The harmonic functions $H_{0,A}, \  H^A$  are  parameterized as:
\be090
H_0 = 1 + {q_0 \over r} \qquad , \qquad H_A = 1 + {q_A \over r} \qquad ,
\qquad H^A = 1 + {p^A \over r} \ .
\ee
Microscopically, this solution corresponds to the case of an
intersection of three $M5$-branes. ($p^A$ is the magnetic charge of  a 5-brane
wrapped around the $A^{\rm th}$ 4-cycle of the Calabi-Yau three-fold
($CY_3$).)  Momentum modes traveling along the common intersection
produce the electric charge $q_0$. (In M-theory compactified on
$CY_3\times S^1$, it is the momentum along $S^1$.) The electric
charge contributions ($q_A$) provide a  shift in ${\tilde H}_0$ (see eq.
(\ref{074})),  which are   due to the  membrane excitations 
with their  flux contribution along   2-cycle $\times  S^1$
of the 5-branes \cite{060},  thus 
  producing additional winding states on $S^1$. 

The  entropy   of the above  BPS-saturated solution is given by:
\be100
{\cal S} = 2\pi \sqrt{\tilde q_0 \, d_{ABC} p^A p^B p^C}\ , 
\ee
where $\tilde q_0$  (determined in terms of
$q_0$ and  $q_A, \ p^A$'s) is  an ``effective'' electric charge 
obtained from  $\tilde H_0$ in (\ref{074}).
The above  entropy has been recently determined
microscopically   in terms of a
2-dimensional $(0,4)$ sigma model whose target  space includes the 5-brane
moduli space 
\cite{060}. (In  \cite{060}  the  microscopic origin of  the 
 loop corrections to the above  tree level entropy  (\ref{100}) have  also been
 discussed.)

\section {Near-BPS-Saturated  Black Holes}

Our aim is to find the entropy  of the near-BPS-saturated solutions. 
Its form should be  determined in terms of  the 
 BPS-saturated  entropy  and 
corrections, parameterized   by the non-extremality parameter $\mu$, 
which quantified a
deviation from the BPS-saturated limit. 

While the BPS-saturated solutions are 
obtained by solving the Killing spinor equations (coupled
first order differential equations),  the non-extreme solutions
solve  general  Euler-Lagrange equations of motion 
(coupled second-order differential equations), a much more formidable task.
However,   since the (bosonic)
action and the equations of motions, including their symmetries,
remain the same, we still assume that the symplectic structure remains
intact. 

\subsection{Equations of Motion}

The gauge field equations and the Bianchi identities follow from
(\ref{010}) and  can be written as
\be107
dF^I = dG_I = 0 \ .
\ee
In order to  solve these equations, it is sufficient to consider only the
spatial components ($\{m,n\}$);  $F_{0m}^I$ can be obtained from $G_{I\, mn}$
by using (\ref{020}). We take the following Ans\" atze: 
\be140
G_{I\, mn} = \tilde \chi_{I}^{\ J}\, \epsilon_{mnp}\partial_p H_{J} \qquad ,
\qquad 
F^I_{mn} = \chi^I_{\ J} \, \epsilon_{mnp}\partial_p H^J \ ,
\ee
where the matrices $\chi$ and $\tilde \chi$ are in general functions of
the radius $r$ and will determined later. The
Ans\" atze (\ref{140})  can  be viewed as a  (local) superposition of the 
extreme solution,
however, note, that  $\chi$ and $\tilde \chi$ are in general  functions of $r$.

For the discussion of the Einstein equations we use the
relation (\ref{020}) and write the gauge field part in terms of
the magnetic fields. With this replacement the gauge field part
in the action (\ref{010})  becomes
\be106 
\int \sqrt{-g} F^I_{\mu\nu} ({^{\star}G_{I}})^{\mu\nu} = 
\int \sqrt{-g} \left( {\rm Im}{\cal N}_{IJ} F^I_{\mu\nu} F^{J\, \mu\nu} + 
{\rm Re}{\cal N}_{IJ} F^I_{\mu\nu} (^{\star}F^{J})^{\mu\nu} \right) \ .
\ee
The second part in this expression is topological, i.e.\ it 
is independent of the metric and does not contribute to the
Einstein equations. The first part  yields the energy momentum tensor and 
we use again (\ref{020})
to replace the electric components by the magnetic ones , i.e.
\be141
F^{I\, 0m} = - {1 \over 2} {\rm Im}{\cal N}^{IJ} \epsilon^{0mnp}
\left(G_{J\, np} - {\rm Re}{\cal N}_{JK} F^{K}_{np} \right) \equiv - 
{1 \over 2} {\rm Im}{\cal N}^{IJ} \epsilon^{0mnp} {\bf G}_{J\, np} \ ,
\ee
with ${\rm Im}{\cal N}^{IJ} = ({\rm Im}{\cal N}_{IJ})^{-1}$.

Thus the  Einstein equations take the form 
\be105
 \ba{l}
R_{00}  = - {1 \over 8} \left(F^2 + {\bf G}^2 \right) g_{00} \\
R_{mn} = 2 g_{A\bar B} (\partial_{m} z^A \partial_{n} \bar z^B ) 
+ {1 \over 2}\left( F^2_{mn} + {\bf G}^2_{mn} - {1\over 4}
(F^2 + {\bf G}^2) g_{mn} \right) \ , 
\ea
\ee 
where $F^2_{mn} = {\rm Im}{\cal N}_{IJ} (F^{I})_{ml} (F^{J})^{\ l}_{n}$
and ${\bf G}^2_{mn} = {\rm Im}{\cal N}^{IJ} ({\bf G}_{I})_{\, ml} 
({\bf G}_J)^{\ l}_{n}$. Insertion of  the gauge fields (\ref{140})
yields
\be108
 \ba{l}
R_0^{\ 0}  = - {1 \over 4} e^{4U} \left(
{\rm Im}{\cal N}_{IJ} \, \chi^I_{\ K} \chi^{J}_{\ L}
\partial_m H^K \partial_m H^L + 
{\rm Im}{\cal N}^{IJ} \tilde \chi_I^{\ K} \tilde \chi_{J}^{\ L} \, 
\partial_m {\cal H}_K \partial_m {\cal H}_L \right) \ , 
 \\
R_{\theta}^{\ \theta} = R_{\phi}^{\ \phi} = - R_0^{\ 0}\ , \\
R_{r}^{\ r} = 2 g_{A\bar B} (\partial_{r} z^A \partial^{r} \bar z^B ) 
+ R_0^{\ 0}\ ,
\ea
\ee 
where 
\be111
\partial_m {\cal H}_I = \partial_m H_I - {\rm Re}{\cal N}_{IK}
\partial_m H^K \ .
\ee
(See (\ref{141}) for the definition of ${\bf G}_{mn}$.)
 
In order to solve  eqs. (\ref{108}), we take the  the following metric
 Ansatz:
\be110
ds^2 = - e^{2U} f  dt^2 + e^{-2U} 
\left( {dr^2 \over f} + r^2 d \Omega \right)\ .
\ee
Consequently, we find the following identity for the 
$R_0^{\ 0} + R_{\theta}^{\ \theta}$  combination:
\be112
R_0^{\ 0} + R_{\theta}^{\ \theta} = {1 \over 2} e^{2U} \left[
f'' + {4 \over r} f' - {2(1 - f) \over r^2} \right] \ .
\ee
It follows from  (\ref{108}) that this expression has to vanish
and we obtain
\be114
f = 1 - { \mu \over r} \ .
\ee
Here $\mu$ is the non-extremality parameter which parameterizes a
deviation from the BPS-saturated limit.

Inserting  the explicit form (\ref{114})  for $f$ into the metric Ansatz
(\ref{110})
 we find for the  Ricci-tensor 
components 
\be116
\ba{l}
R_{0}^{\ 0} = - R_{\theta}^{\ \theta} = - R_{\phi}^{\ \phi} 
= e^{2U} \left[ f \partial^2 U + \partial_m f  \partial_m U \right] \ , \\
R_{r}^{\ r} = e^{2U} \left[ -f (\partial^2 U - 2 \partial_m U 
\partial_m U) + \partial_m f \partial_m U \right] \ .
\ea
\ee
where $\partial^2 U = U'' + {2 \over r} U'$ is the flat space Laplacian.
The  symmetries between the Ricci tensor components 
are in agreement with eq.\  (\ref{108}).

Finally, the scalar field equation is given by
\be109
\ba{l}
{4 \over \sqrt{g}} \partial_{\mu} (\sqrt{g} g^{\mu\nu} g_{A\bar B} 
\partial_{\nu} \bar z^{B}) - 2 (\partial_{A} g_{B\bar C}) \partial z^{B} \,
\partial \bar z^{C}   \\ \qquad - {1 \over 4} \left[
(\partial_A {\rm Im}{\cal N}_{IJ}) F^{I}_{mn} F^{J\, mn} +
(\partial_A {\rm Im}{\cal N}^{IJ}) {\bf G}_{I\, mn} {\bf G}_J^{mn} \right]
= 0\ ,
\ea
\ee
where $\partial_A = {\partial \over \partial z^A}$.

Since we assume that the symplectic structure remains intact,
$e^{-2U}$ in the metric is still given by the   structure of the 
K\"ahler potential.
However, now we allow in $e^{-2U}$ the replacement of harmonic
functions $(H^A, H_I)$ by {\it general functions} $(\bar H^A , \bar
H_I)$. Thus, for the model with the prepotential (\ref{prep}) 
and the assumption that $\bar H^0 = 0$ the
Ansatz is of the form:
\be120
e^{-2U} = \sqrt{{\hat H}_0 d_{ABC} \bar H^A \bar H^B \bar H^C} \quad , 
\quad z^A = {1 \over 6} D^{AB} \bar H_B  -i \hat H_0 \bar H^A e^{2U} \ .
\ee
where $\hat H_0 = \bar H_0 + {1 \over 12} D^{AB}\bar H_B \bar H_A$.
In terms of these functions the Einstein equations 
(\ref{108}) reduce to one single equation
\be160
\ba{r}
- 2 \, \partial_m f \partial_m e^{-2U} =
\mbox{Im}{\cal N}^{IJ}
\left( f \partial_m {\bar {\cal H}_I} \partial_m {\bar{\cal{H}}_J} - 
\tilde \chi_{I}^{\ K}  \tilde \chi_{J}^{\ L} \partial_m {{\cal H}_K} 
\partial_m {\cal H}_L \right) - 8  \, {\rm Re} X^I \, \partial^2 \bar H_I  \\
+ \mbox{Im}{\cal N}_{IJ}
\left( f \partial_m \bar H^I \partial_m \bar H^J - \chi^{I}_{\ L} 
\chi^{J}_{\ K} \partial_m H^L \partial_m H^K \right) + 8\,
{\rm Re} F_I \, \partial^2 \bar H^I\ , 
\ea
\ee
where we have used the known expression for $\partial^2 U$ (from the extreme
solution). We can also simplify the scalar field equation
\be166
\ba{l}
-8\,  e^{-2U} \partial_m f \, g_{A\bar B} \partial_m z^{\bar B} = \hfill
\\ \qquad 
(\partial_A \mbox{Im}{\cal N}^{IJ})
\left( f \partial_m {\bar {\cal H}_I} \partial_m {\bar{\cal{H}}_J} - 
\tilde \chi_{I}^{\ K}  \tilde \chi_{J}^{\ L} \partial_m {{\cal H}_K} 
\partial_m {\cal H}_L \right) + {\cal O}(\partial^2 \bar H_I)  \\
\qquad + (\partial_A \mbox{Im}{\cal N}_{IJ})
\left( f \partial_m \bar H^I \partial_m \bar H^J - \chi^{I}_{\ L} 
\chi^{J}_{\ K} \partial_m H^L \partial_m H^K \right) + 
{\cal O}(\partial^2 \bar H^I)\ . 
\ea
\ee
We omitted here the explicit expressions $\sim \partial^2 \bar H$,
because they do not have a simple form and we do not need them later.
This structure of the Einstein and scalar equation suggests that one
can still try to find a solution for $( \bar H^I , \bar H_I)$ in terms
of harmonic function, at least in the neighborhood of the horizon.

\subsection{The near-extreme solution}


In order to obtain the complete solution one has to solve  explicitly
 both the Einstein
equations as well as the scalar
field equation. However, in order to obtain the  entropy for 
near-BPS-saturated black holes, it is sufficient to consider  the behavior of
the solution  in the near-region of  the outer-horizon. 
We will assume that  in this region 
  the general functions $(\bar H^I , \bar H_I)$ of 
    near-BPS-saturated solution  
can again be approximated by harmonic functions. Hence, we take
\be163
\bar H^I  = {\bar p^I \over r} + \bar h^A  + {\cal O}(\mu) \ ,
\ee
and a similar expression for $\bar H_I$. In order to solve the
Einstein equations (\ref{160}) we consider the expansion
\be121
\partial_m \bar H^I = \Omega^I_{\ J} \partial_m H^J
\qquad , \qquad 
\partial_m \bar {\cal H}_I = \tilde \Omega_I^{\ J} \partial_m {\cal H}_J\ ,
\ee
where $\Omega$ and $\tilde \Omega$ will be fixed later.
Using the definition of K\"ahler potential (\ref{030}) we find for
 $\partial_m e^{-2U}$ 
\be162
\ba{rcl}
-2 \partial_m e^{-2U} &=& U^I \, \partial_m \bar {\cal H}_I + U_I \,
\partial_m  \bar H^I \\
&=& -4 \,{\rm Re}X^I \, \partial_m \bar {\cal H}_I 
+ 4\, ({\rm Re}F_I - {\rm Re}X^J {\rm Re}{\cal N}_{JI}) \, 
\partial_m \bar H^I\ .
\ea
\ee
It is straightforward to read off the coefficients $(U^I , U_I)$
for the Ansatz given in (\ref{120}).

Before we discuss the general solution, let us first recall the
non-extreme solution for the toroidal compactification, which
corresponds to the case where from the intersection numbers $d_{ABC}$
($A,B,C=1,2,3$) only  $d_{123}={1\over 6}$ is
non-vanishing. If one furthermore assumes that all
axions vanish, i.e. \ if $H_A=0$, the matrices $\Omega$ and $\chi$ are
known \cite{040}
\be180
\ba{lll}
\bar H^A = (\Omega_0)^A_{\  B} H^B = \tanh\beta^{(A)} \delta^{A}_{\ B} \, 
H^{B} &,&\bar H_0 = \tanh\beta^{(0)} H_0 \quad , \quad
\chi = {\bf 1} \ ,
\ea
\ee
 the charges are given by
\be181
\ba{lll}
\bar p^A = \mu \sinh^2\beta^{(A)} &,&
p^A = \mu \sinh\beta^{(A)} \cosh\beta^{(A)} \ , \\
\bar q_0 = \mu \sinh^2\beta_{(0)} &,&
q_0 = \mu \sinh\beta_{(0)} \cosh\beta_{(0)} \ ,
\ea
\ee
and the electric gauge field  are specified by  $F_{0r}^0 = \mbox{Im}{\cal
N}^{00} (^{\star}G_0)_{0r} = {q_0 /(\bar H_0^2 r^2)}$ (see
(\ref{020})).

In order to obtain the above ``toroidal'' solution, it was 
crucial that the matrix
Im${\cal N}_{AB}$ was {\it diagonal}.  In general this is not the case,
however, in the near-horizon region of a general near-BPS-saturated solution
 one can still
diagonalize Im${\cal N}_{IJ}$ by using the constant matrices $\Omega$
and $\chi$ and ${\rm Im}{\cal N}^{IJ}$ by $\tilde \Omega$ and $\tilde
\chi$. We take
\be208
\Omega = \chi \Omega_0  \quad , \quad \tilde \Omega = \tilde \chi
\tilde \Omega_0 
\ee
with  $\chi$ and $\tilde \chi$ defined by
\be209
\ba{c}
(\chi^T {\rm Im}{\cal N}^{-1} \chi)^{IJ} = {\rm diag}(\lambda^{(0)}\ ,
\ \lambda^{(1)} \ , \  ..) \quad , \quad \lambda^{(I)} = 
U^I/\bar {\cal H}_I \ , \\ 
(\tilde \chi^T {\rm Im}{\cal N} \tilde \chi)_{IJ} = 
{\rm diag}(\rho^{(0)} \ , \, \rho^{(1)} \ , \  ..) 
\quad , \quad \rho^{(I)} = U_I/\bar H^I \ , \\
\ea
\ee
where ($U_I , U^I$) were introduced in  (\ref{162}).
Note, that $\lambda^{(I)}$ and $\rho^{(I)}$ are not eigenvalues  of 
${\rm Im}{\cal N}^{-1}$ and ${\rm Im}{\cal N}$,
but they are proportional to them. Thus, the harmonic functions
are given by
\be210
\ba{lll}
\bar {\cal H}_I =  \tilde \chi_I^{\ J} 
\left(1 + {\mu \sinh^2\beta_{(J)} \over r} \right) 
&,&  {\cal H}_I =  1 + {\mu \sinh\beta_{(I)} \cosh\beta_{(I)}
\over r}  \ , \\
\bar H^I = \chi^I_{\ J} \left( 1 + {\mu \sinh^2\beta^{(J)} \over r} \right)
&,&  H^I = 1 + {\mu \sinh\beta^{(J)} \cosh\beta^{(J)}
\over r}  \ .
\ea
\ee

Finally, we should express the parameter in the metric ($\bar q_I ,
\bar p^I$) by physical charges $\tilde q_I = \tilde \chi_I^{\ J} q_J$ and
$\tilde p^I = \chi_{\ J}^I p^J$, obtained from (\ref{140}) and find
\be220
\bar p^I = \Omega^I_{\ J} p^J =  
(\chi \Omega_0 \chi^{-1})^{I}_{\ J} \, \tilde p^J \qquad , \qquad
\bar q_I = \tilde \Omega_I^{\ J} q_J = 
(\tilde \chi \tilde \Omega_0 \tilde \chi^{-1})_I^{\ J} \, \tilde q_J \ .
\ee
where $\Omega_0$ and $\tilde \Omega_0$ are given by
\be222
\Omega_0 = {\rm diag}(\, \tanh \beta^{(1)} , \tanh \beta^{(2)} , ... \, ) 
\qquad , \qquad
\tilde \Omega_0 = {\rm diag}(\, \tanh \beta_{(1)} , \tanh \beta_{(2)} , ... \,
)
\ee
On the horizon $r= \mu$, the harmonic functions $(\bar H_I,\ {\bar H}^I)$ 
become
\be224
\mu \bar {\cal H}_I \Big|_{horizon} =  \tilde \chi_I^{\ J} 
\, \mu \cosh^2\beta_{(J)}  \qquad , \qquad
\mu \bar H^I \Big|_{horizon} = \chi^I_{\ J} \, \mu \cosh^2\beta^{(J)}\ .
\ee
Then the  Hawking entropy for  the  model specified by  (\ref{120}) is of the
form
\be230
{\cal S} = 2\pi \sqrt{\mu \hat H_0 \, d_{ABC} \, \mu\bar H^A \, 
\mu\bar H^B \, \mu\bar H^C}
\Big|_{horizon}\ , 
\ee
and the Hawking temperature becomes
\be240
T = {\mu \over 4 \pi \sqrt{\mu \hat H_0 \, d_{ABC} \, \mu\bar H^A 
\, \mu\bar H^B\,  \mu\bar H^C}
|_{horizon} } \ .
\ee
Replacing the charges ($\bar q_0 , \bar p^A$) by the physical charges
($\tilde q_0 , \tilde p^A$) from eq.\ (\ref{220}) we obtain 
corrections to the  extremal (BPS-saturated) entropy.


\section{Microscopic Interpretation of the Near-Extreme Black Hole Entropy}


A microscopic interpretation of the classical entropy in the
BPS-saturated limit (\ref{100}) was given in \cite{060} and
\cite{026}. Let us shortly summarize the main points.  The model at
hand appears as an intersection of three M5-branes, which intersect
over a common string  and 
momentum modes travelling along this common string. 
The momentum modes are parameterized by the electric charge
$q_0$ and the 5-branes are related to the magnetic charges $p^A$. In
addition, the model described by (\ref{074}) contains further electric
charges $q_A$, which correspond to the membranes, specified by the
5-brane world-volume excitations of the self-dual antisymmetric
tensor.  For the common string these modes appear as winding states
shifting the vacuum energy and momentum. So, the microscopic model can
be described by a (1+1)-dimensional sigma model for the common string
which has in the BPS-saturated limit a (4,0) supersymmetry.  As a
consequence of the world-sheet supersymmetry, all the
momentum modes are chiral, e.g., only the left-moving excitations are
present~\footnote{There is a subtlety associated with the membrane
modes, which can possess one right-moving mode coming from the
two-form on the membrane, and which are paired with the three
translational zero modes and four goldstinos in a $(4,0)$ multiplet.
}. In addition, the 4-cycles of the Calabi-Yau
three-folds, around which the 5-branes
are wrapped, are holomorphic. 
Both  of these properties  are lost in the 
near-BPS-saturated  case, which we will discuss now.

In comparison to the black hole solutions of toroidally compactified string
theory ($N=4,8$ string vacua),  the near-extreme black holes of $N=2$ string
vacua possess one
obvious modification: the charges  undergo a rotation, which is 
caused by the non-diagonal gauge coupling matrix ${\cal N}_{IJ}$ and
is expressed by the mixing matrices $\chi$ and $\tilde \chi$ in
(\ref{224}). However,  barring this rotation of charges, 
the physical  microscopic  interpretation 
of the solution 
resembles 
the
toroidal case. First, the physical charges, which are defined as
integral over the gauge fields (\ref{140}) with the harmonic function
defined in (\ref{210}), split into two contributions
\be242
 \sim \mu \, \sinh\beta \cosh\beta = \mu \, (e^{2\beta} - e^{-2\beta}) = 
   Q - \bar Q\ , 
\ee
where the first part coincides with the extreme charge and the second
part is the near-extreme perturbation. In \cite{020} the second part
has also been interpreted as a small contribution with the opposite
charge (e.g., \ anti-brane).  This splitting takes place for all 
charges: (i) for
the magnetic charges $p^A$ it implies, that the 5-brane wraps also an
anti-holomorphic part of the 4-cycle and in the BPS-saturated limit 
($\mu\to 0$,
$\beta^A\to \infty$, with $p^A \sim \mu e^{2\beta^A}$ finite) the
anti-holomorphic parts vanish, (ii)  for the momentum
modes (associated with the electric charge $q_0$) this result simply
implies that there are  left- as well as right-moving momentum modes along
$S^1$, and finally (iii), 
the electric charges ($\sim q_A$) of the 
membranes  with non-trivial fluxes through  a 2-cycles $\times S^1$, yield 
winding as well as ``anti-winding'' 
states on  $S^1$.

Although this splitting is valid for all the three types of  charges,
 in the  subsequent discussion of
the microscopic entropy we concentrate on the momentum part associated
with $q_0$, i.e.   we will consider the microscopic in a 
dilute gas
approximation.  This means we will take large magnetic boosts so that
\be250
\Omega_0 \simeq {\bf 1} \qquad {\rm or} \qquad  \bar p^I \simeq \tilde p^I
\simeq p^I \ .
\ee
For the sake of simplicity 
 we consider the axion-free case with $\bar
H_A =0$. Because in this case ${\cal N}_{0A} =0$ also $\chi^0_{\ A} =
\tilde \chi_0^{\ A} =0$ and hence the 0-components factorize.   In this case 
the form of the classical entropy takes the form:
\be260
\ba{l}
{\cal S} = {\cal S}_L + {\cal S}_R = 2 \pi 
(\sqrt{(N_L ) c_L/6}+\sqrt{( N_R) c_R/6})\ ,
\\ {\cal S}_L =
 2\pi \sqrt{\tilde \chi_0^{\ 0} \mu \, e^{2\beta_{(0)}}} \;
\sqrt{d_{ABC} p^A p^B p^C } = 2 \pi \sqrt{N_L c_L/6}
\\
{\cal S}_R = 2\pi \sqrt{\tilde \chi_0^{\ 0} \mu \, 
e^{- 2\beta_{(0)}}} \; \sqrt{ d_{ABC} p^A p^B p^C } =
2 \pi \sqrt{N_R c_R/6}\ .
\ea
\ee
i.e.\ both, the left and right-moving modes feel the the
same central  charge  ($c_L=c_R$), 
since 
in the dilute gas approximation we consider the
5-branes wrapped around supersymmetric (holomorphic) 4-cycles (see \cite{060}).
 The momentum modes $N_{L,\, R}$ are respectively the  left- and right-moving 
momenta along $S^1$.


We would  like to conclude with a few remarks.
\begin{itemize}
\item 
{\it Duality invariant structure of the thermodynamic quantities}
The result  for the entropy and the temperature of
  the near-BPS-saturated solution  is cast in a 
 duality invariant form, since 
 by construction the K\"ahler potential is duality invariant.

\item 
{\it General non-extreme solutions}
The equations of  motion  may have 
an explicit solution for a  (large) class of non-extreme solutions.
In particular, there is a possibility of solving
explicitly the scalar equation (\ref{166}) simultaneously with the Einstein
equations (\ref{160}) and 
a large class of the
solutions can be expressed in terms of the harmonic functions, only.
  
\item 
{\it Rotating non-extreme solutions}
Another interesting avenue  to pursue  are  rotating solutions. 
As a first step, of most interest are the near-BPS-saturated 
rotating solutions.
(In the BPS-saturated limit, only static four-dimensional solutions have 
regular horizons.)  From the point of view of the microscopic interpretation 
one expects the angular momentum quanta 
to be identified  with the $U(1)_R\in SU(2)_R$ world-sheet
charges  of  the  right-moving  modes. (The  right-moving 
 sector of the 2-dimensional
sigma model possesses $N=4$ world-sheet supersymmetry  with 
the $SU(2)_R$  world-sheet current algebra.)

The ultimate goal would be to obtain 
 non-extreme rotating solutions (both
in four- and five-dimensions), and hopefully arrive at the same suggestive
structure of the thermodynamic quantities, as in the case of black holes of
$N=4,8$ superstring vacua \cite{140}; in the latter case these quantities 
 split into  two,  suggesting a microscopic interpretation in 
  terms of  effective string
degrees of freedom.

\item {\it Calabi-Yau phase transitions} 
 The structure
of this near-extreme solution has a remarkable property,  that the 
Bekenstein-Hawking
entropy and the Hawking temperature depend  on the gauge coupling
matrix ${\cal N}_{IJ}$. As long as it is regular, a non-vanishing
$\mu$ can be treated as a small fluctuation around the extremal
limit. However at points where ${\cal N}_{IJ}$ or ${\cal N}^{IJ}$
becomes singular, already small fluctuation could cause an infinite
Hawking temperature making the black hole unstable.  These are the
interesting points where Calabi-Yau phase transition can occur.  
The corresponding BPS-saturated solutions have been investigated in
\cite{070} and they remain finite as long as one keeps at least two
charges. It would be very interesting to investigate
these  properties further.

\end{itemize}

\bigskip \bigskip

{\bf Acknowledgments}

The work is supported by the Deutsche Forschungsgemeinschaft (DFG)
(K.B.,W.A.S.) and by the U.S. Department of energy grant
DE-FG02-95ER40893 (M.C.).  We would like to thank F. Larsen for
discussions and S. Mathur for correspondence. K.B. would like to thank
the Department of Physics ans Astronomy of the University of
Pennsylvania for the hospitality, where part of this work has been
done.  W.A.S, would like to thank the Institute of Theoretical
physics, ETH, Z\"urich and the Particle Theory group at The
Rockefeller University, where part of this work was done, for
hospitality.


%
%

\end{document}